\documentclass[aps,amsmath,twocolumn]{revtex4}
\usepackage{graphicx}
\usepackage{subfig}
\usepackage{relsize}

\def\flip#1{\reflectbox{#1}}

\begin{document}


\title{Pi in the Sky}

\author{Ali Frolop} \email{afrolop@phas.ubc.ca}
\author{Douglas Scott} \email{docslugtoast@phas.ubc.ca}
\affiliation{Dept.\ of Physics \& Astronomy,
 University of British Columbia, Vancouver, Canada}

\date{1st April 2016}

\begin{abstract}
Deviations of the observed cosmic microwave background (CMB) from the
standard model, known as `anomalies', are obviously highly
significant and deserve to be pursued more aggressively in order to discover
the physical phenomena underlying them.  Through intensive investigation
we have discovered that there are equally surprising features in the digits
of the number $\pi$, and moreover there is a remarkable
correspondence between each type
of peculiarity in the digits of $\pi$ and the anomalies in the CMB.
Putting aside the unreasonable possibility that these are just the sort of
flukes that appear when one looks hard enough, the only conceivable
conclusion is that, however the CMB anomalies were created,
a similar process imprinted patterns in the digits of $\pi$.
\end{abstract}

\maketitle

\noindent
\section{Introduction}
The standard cosmological model successfully describes a wide range of
observational phenomena using just six parameters in the context of a framework
that requires only a handful of basic assumptions \cite{mnemonics}.
The `$\Lambda$CDM'
model is so successful in fact that attention has focussed on deviations from
this simple picture.  Just like with particle
physics, one can gauge the maturity of the field through the fact that
activity switches from establishing the validity of the framework
to searching for physics `beyond the standard model'.

The most precise cosmological data come from careful measurements
of the anisotropies in the cosmic microwave background (CMB \cite{cmb}),
as surveyed by the {\it COBE} \cite{COBE}, {\it WMAP\/} \cite{WMAP}
and {\it Planck\/} \cite{PlanckI} satellites, as well as with
a suite of ground-based and balloon-borne experiments.  {\it COBE\/}
helped establish $\Lambda$CDM as the standard cosmological model, and 
for many people the biggest news from both {\it WMAP\/} and {\it Planck\/} was
that there is no news, i.e.\
the standard model continues to be a good fit, even as the precision
has improved dramatically.  Since the days of {\it COBE\/} (e.g.\
Ref.~\cite{nulltests}) there has been intensive
investigation into deviations from Gaussianity or the breaking of statistical
isotropy on large scales, motivated by the fact that they could be
smoking guns for new physics in the
early Universe.  This process has continued to the present day,
with a huge number of studies
searching for blemishes of various forms in the CMB sky, which might be
evidence for chinks in the armour of the standard model \cite{sesame}.  Now
it seems like this search has become the main industry in cosmology.

Hundreds of papers have been written on this topic.  We cannot possibly
refer to all of them here, but many can be found in the reference lists of
several overviews of the subject \cite{anomalies,WMAPA,IandS}.
In the CMB sky, such deviations from perfection are often referred to as
`anomalies', and there are several distinct features of this sort that have
been identified.  Sarkar et al.\ \cite{improbable0} pointed out that there are
at least two unrelated kinds of anomalies and hence one can take the
product of the chance probabilities to determine how unlikely our observed
Universe is in the standard model.  Recently Schwarz et al.\ \cite{improbable1}
have argued that there are at least {\it three\/} distinct kinds
of anomaly, and since each of them separately has a probability of
${\sim}\,10^{-2}$ in the standard model, then our CMB sky is unlikely at the
roughly 1-in-a-million level.  Similar conclusions were also made in another
recent paper by Melia \cite{improbable2}.

In fact considerably more than three kinds of anomaly have already been
identified, and no doubt more remain to be uncovered.  They include:
the Cold Spot; the low-$\ell$ deficit; the $\ell\,{\simeq}\,20$--30 dip;
hemispheric asymmetry; low variance;
dipole modulation; odd/even multipole asymmetry; and other specific features,
such as rings, stripes, wiggles, bites and even letters
\cite{crazy,naysayers}.  When considered together, the
combined probability is vanishingly small that our sky could be a realisation 
of a Gaussian random process within $\Lambda$CDM.

Amazing though these CMB features are, we have discovered that a much more
familiar data set, namely the digits of $\pi$ \cite{flip}, contains equally
astonishing anomalies.  Through the use of advanced analysis methods, using
pattern recognition software and Bayesian search algorithms, we have discovered
many distinct features in $\pi$.
In fact, for each CMB anomaly there appears to be a corresponding
effect in the digits of $\pi$ -- we call these analogous anisotropy anomalies
(or AAAs).  These features in $\pi$ are {\it at least\/} as significant
as the cosmological ones, and given that the CMB anomalies are
patently evidence for new physics, this means that the patterns in $\pi$ must
demonstrate the existence of {\it mathematics\/} beyond the standard model
\cite{max}.

\section{The digits of $\boldmath{\mathlarger{\mathlarger{\mathlarger{\pi}}}}$}
The irrational number $\pi$ \cite{rational}
occurs throughout physics, in anything related to
rotations, waves, vibrations or phases.
It occurs explicitly in Coulomb's law, Kepler's third law,
Einstein's field equations, the Fourier transform, the normalization of a
Gaussian, the reduced Planck constant, etc.
And perhaps most astonishingly, if you divide the circumference
of a circle by its diameter, you get exactly $\pi$!

Unlike other numbers, $\pi$ attracts enormous attention, with contests to
memorise its digits \cite{recite}, and it even has its
own day \cite{piday}.  The number has also inspired a style of writing
called `pilish' (whose origin has been attributed to Sir James Jeans
\cite{pilish}), in which one uses words whose lengths match sequentially
to the digits of $\pi$.  Furthermore it is known that the digits of $\pi$
eventually have to contain messages, and there is speculation that these may
reveal basic truths about the nature of reality \cite{contact}.

\begin{figure*}[!htpb]
\begin{center}
 \subfloat[]{\includegraphics[width=0.575\textwidth]{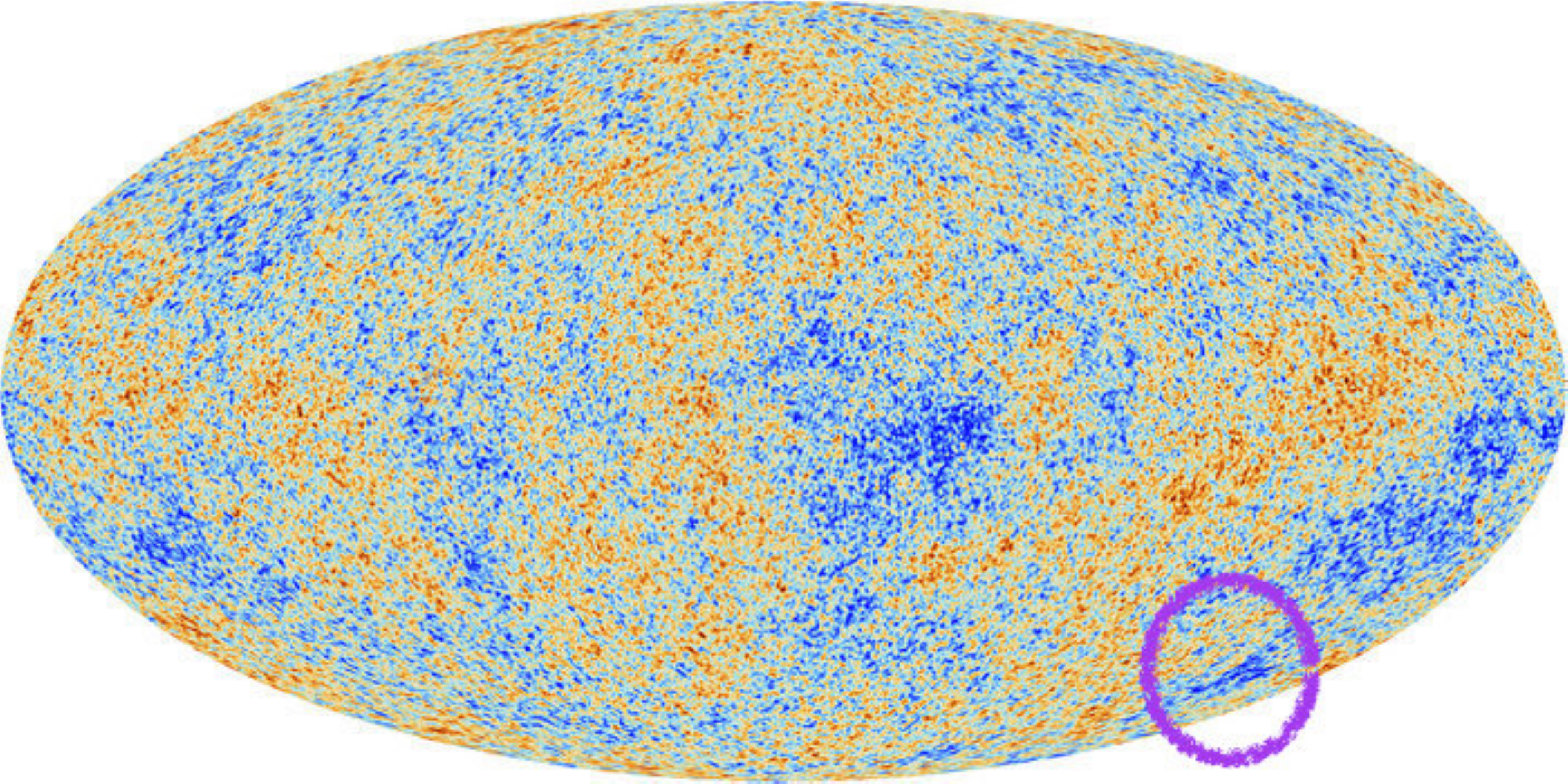}}
 \hfill
 \subfloat[]{\includegraphics[width=0.40\textwidth]{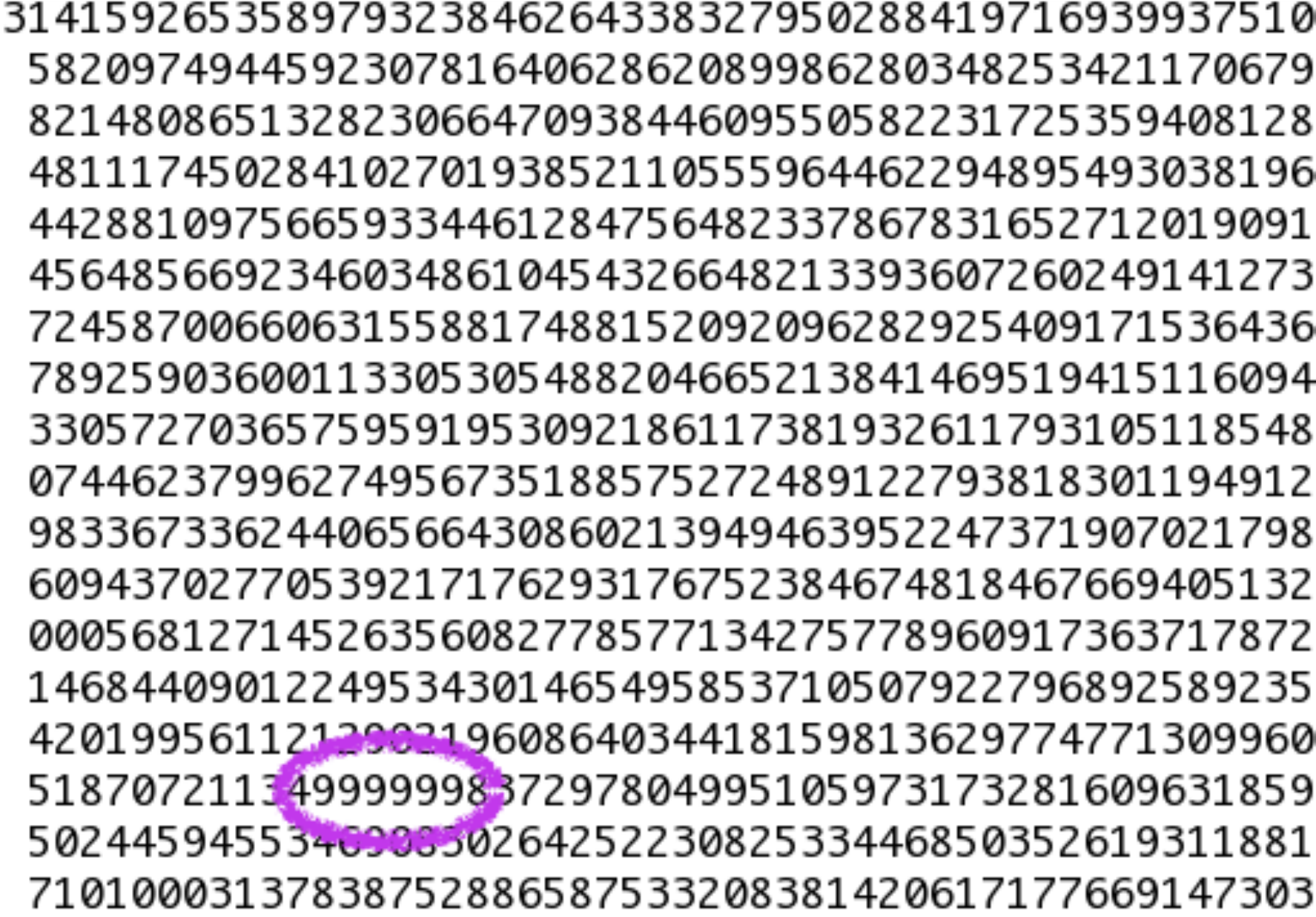}}
\end{center}
\caption{\label{fig:nocoldspot}
(a) Map of the CMB sky \cite{chart} from the {\it Planck\/} satellite
\cite{PlanckI}.  It seems hardly necessary to mark the position of the Cold
Spot, since it stands out so clearly.  (b) The first 900 digits of
$\pi$, showing the early `hot spot', also known as the Feynman point.}
\end{figure*}

By developing novel computational techniques,
we have performed extensive searches for patterns in the digits of $\pi$,
similar to those already carried out for the CMB sky.
For definiteness, in most of the investigations that we describe below,
we focus on the first 10{,}000 digits of $\pi$ \cite{internet}, since this
is similar to the
number of modes probed by today's CMB mapping experiments.  We leave it
to other researchers to extend our study to the trillions of additional
digits that are now known.  Like the CMB sky \cite{future},
we also assume that the digits of $\pi$ are
unchanging, although previous studies have already been carried out to
investigate possible temporal \cite{scherrer} or spatial \cite{knox}
variations of $\pi$.

\section{Anomalies}
\subsection{The Cold Spot}
Let us begin with one of the most widely discussed features in the CMB sky,
namely the `Cold Spot'.  Despite the fact that the power spectrum of the CMB
anisotropies carries around $2000\,\sigma$ worth of information \cite{2dinfo},
the apparently 2--3$\,\sigma$ significance of the Cold Spot is fantastically
important.  A quick glance at Fig.~\ref{fig:nocoldspot}(a) shows just how
prominent this feature is on the sky.  Although not the large blue region
near the centre \cite{pit}, and not actually the coldest place on the
CMB sky \cite{coldest}, the chances of finding a cold region of exactly this
size and shape in Gaussian random skies is quite small.  And the chances of
finding such a spot in precisely this direction is almost vanishingly small.

In the digits of $\pi$ there is an analogous feature, which might be called
a `hot spot', since it involves a cluster of the number `9'.  Specifically,
`9' occurs six times concurrently after the 762nd digit of $\pi$, as shown in
Fig.~\ref{fig:nocoldspot}(b).
This is sometimes called the `Feynman point' \cite{Feynman}, and appears
fantastically earlier in the digits of $\pi$ than one would expect for such a
repeated run.  One can estimate the probability of this AAA
in the following way: the number of
possible outcomes for 762 digits is $10^{762}$; the number of
places the run `999999' could be located within the first 762 digits
is $(762-6)=756$ and the number of possibilities for
the other 756 digits is $10^{756}$; and hence the probability is $756/10^6
\simeq8\times10^{-4}$.

It has also been pointed out \cite{tau} that the digits of $2\pi$ contain
{\it seven\/} consecutive 9s at about the same position as the six 9s in
$\pi$.  This is of course even less likely to occur by chance.

There are many other examples of unlikely runs of digits within $\pi$
\cite{pisearch}.
For example, if we focus on the number `4', we find that the sequence `4444'
occurs much {\it later\/} than expected (in fact it is the last of the sets of
four identical numbers), not until digit 54525.  And amazingly `44444' is also
the last sequence of its type to occur (i.e.\ after `00000', `11111', etc.),
not until digit 808651.  This of course leads us to ask what is special about
4 and about 9 -- what previously hidden mathematical principle could explain
this behaviour?

\begin{figure*}[!htpb]
\begin{center}
 \subfloat[]{\includegraphics[width=0.475\textwidth]{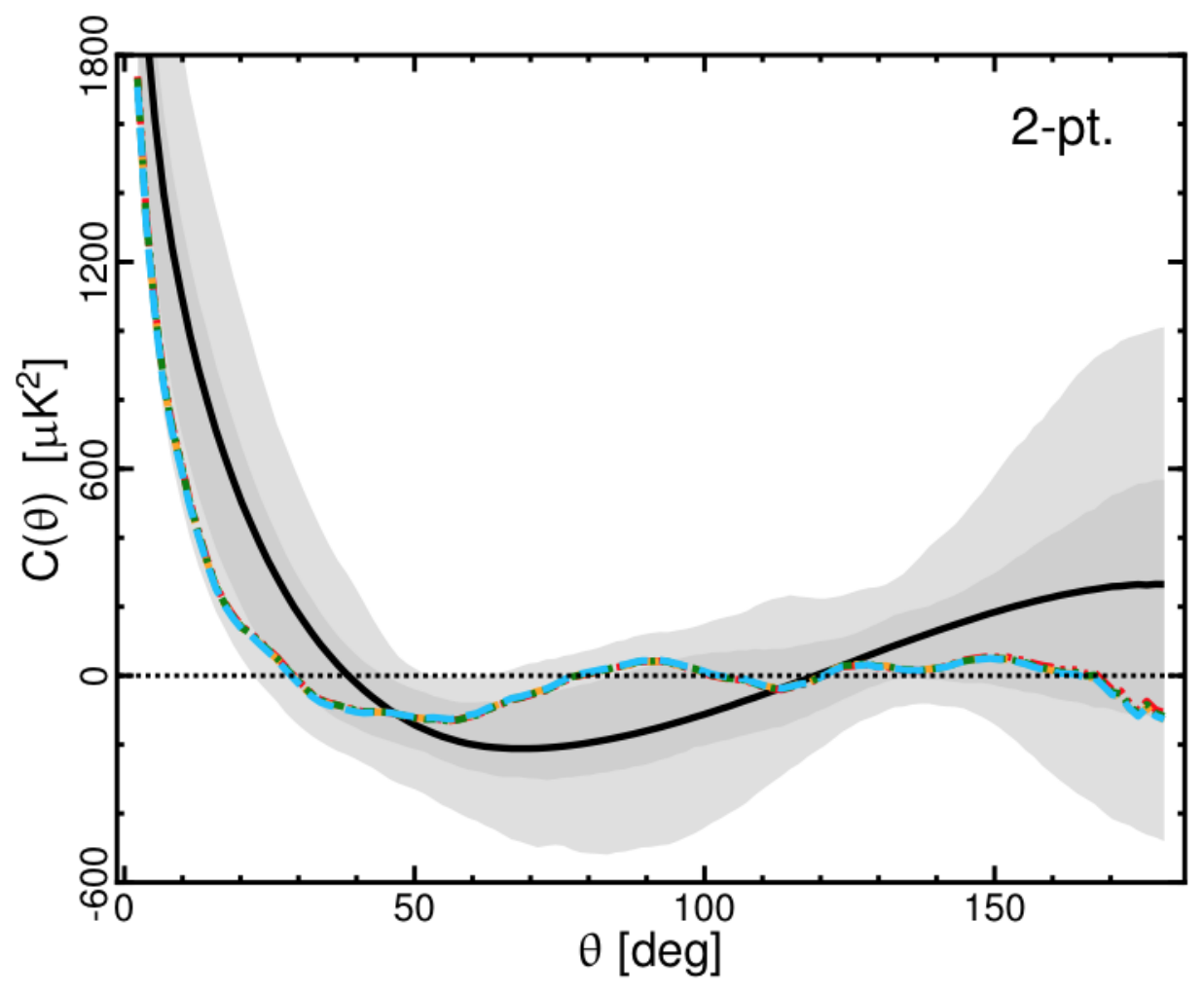}}
 \hfill
 \subfloat[]{\includegraphics[width=0.475\textwidth]{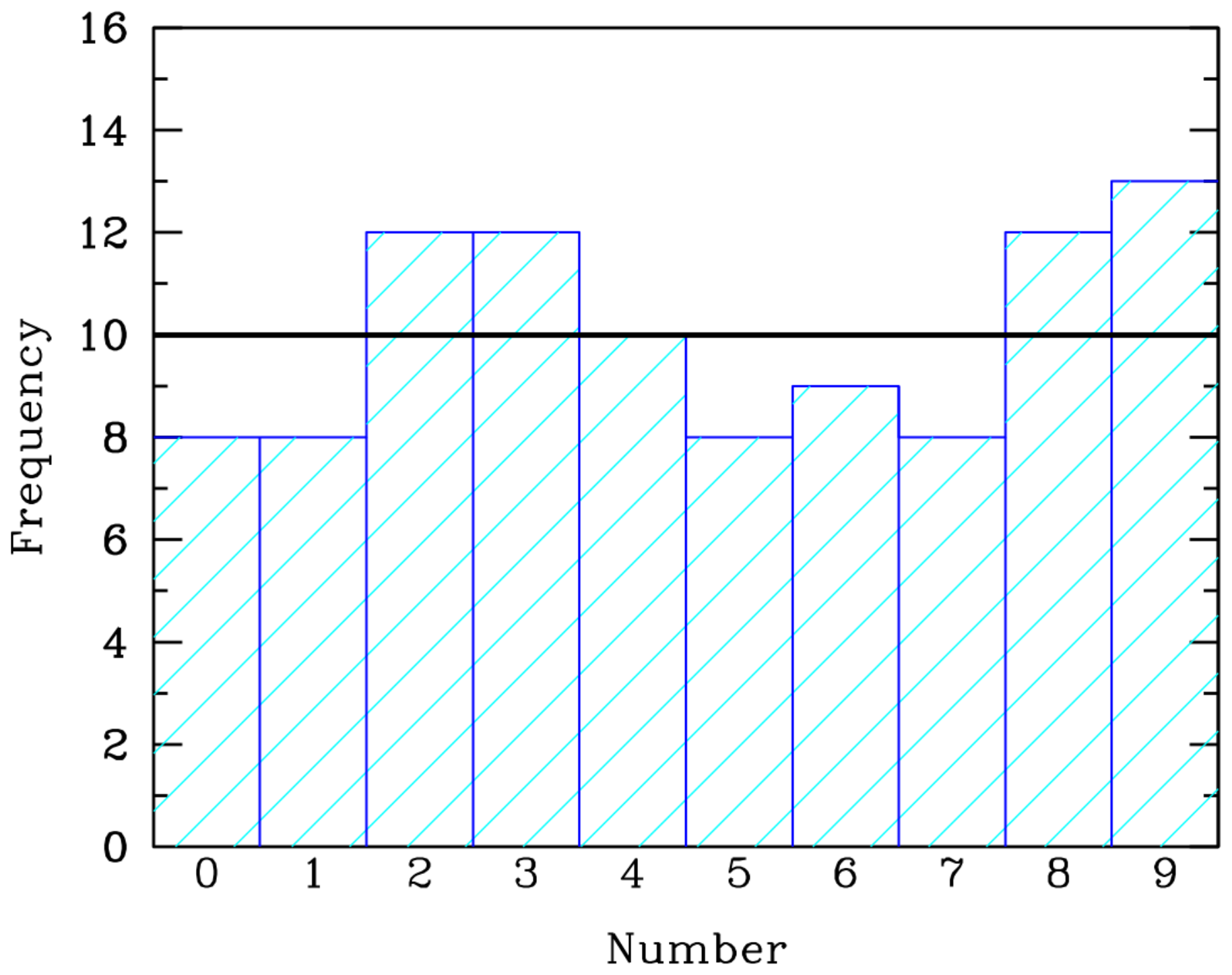}}
\end{center}
\caption{\label{fig:lowvariance} (a) Correlation function for {\it Planck\/}
data, taken from Ref.~\cite{improbable1}.  This is one of the most
striking and well-defined of the CMB anomalies, arising from the fact that the
data are quite close to zero for a fair range of angles, something that must
surely be quite unlikely to find by chance.
(b) An analogy for the low variance of
the CMB for small multipoles is seen in the distribution of numbers in
the first 100 digits of $\pi$.  The
frequency of these digits is remarkably `quiet', i.e.\ all digits occur quite
close to the average number of times (the thick black line), which must also
be very unlikely.}
\end{figure*}

\subsection{Low variance}
The large-scale CMB sky has lower variance than expected, and this
can be examined in several different ways, e.g.\ by estimating the variance
of the temperature field or some filtered power spectrum estimator.
One popular statistical approach is to examine the correlation function of the
CMB, which is shown in Fig.~\ref{fig:lowvariance}(a).  It is very noticeable
that above a precisely defined scale (say about $70^\circ$) the correlation
function is very nearly approximately zero.  And the chances of this are
surely tiny.  This result is so dramatic that the Universe is clearly trying
to tell us something -- perhaps there is an anthropic connection between the
lack of large-scale correlations and our existence?

This deficiency in the variance at low multipoles has an analogy in the
behaviour of the
early digits of $\pi$.  In the first 100 digits, the numbers 0 to 9 all occur
fairly close to the average number of times (which is 10), as shown in
Fig.~\ref{fig:lowvariance}(b).  The biggest
deviation is for the number 9, which occurs 13 times, this being exactly
$1\,\sigma$ from the mean, with no other number occurring with a frequency that
is more than $1\,\sigma$ from the expectation.  This therefore represents a
very `quiet' distribution.  Using Poisson statistics, the chances of picking
100 random digits and finding one digit more than 13 times is 0.220 and the
chances of finding it less than
8 times is 0.136; hence for each of the 10 digits we have
${\rm Prob}(8,9,10,11,12,13)=0.644$ and then the total probability of having
such a quiet distribution is $P=0.644^{10}=0.012$.  It is clear that the
early digits of $\pi$ are already telling us something through the fact that
none of these integers wants to stand out.

In fact there are indications that this low variance continues further, and
we can determine that the expected number for $n$ digits is $np$, with $p=0.1$,
and the variance is $\sqrt{np(1-p)}$.  In the first
1{,}000 digits only one number (`1' in this case) occurs more than
$1\,\sigma$ from the expected frequency, and a similar pattern is even seen
out to 10{,}000{,}000 digits, where `4' is the only number occurring more than
$1\,\sigma$ from the expected fraction.

\subsection{Low-$\ell$ deficit}
\begin{figure*}[!htpb]
\begin{center}
 \subfloat[]{\includegraphics[width=0.525\textwidth]{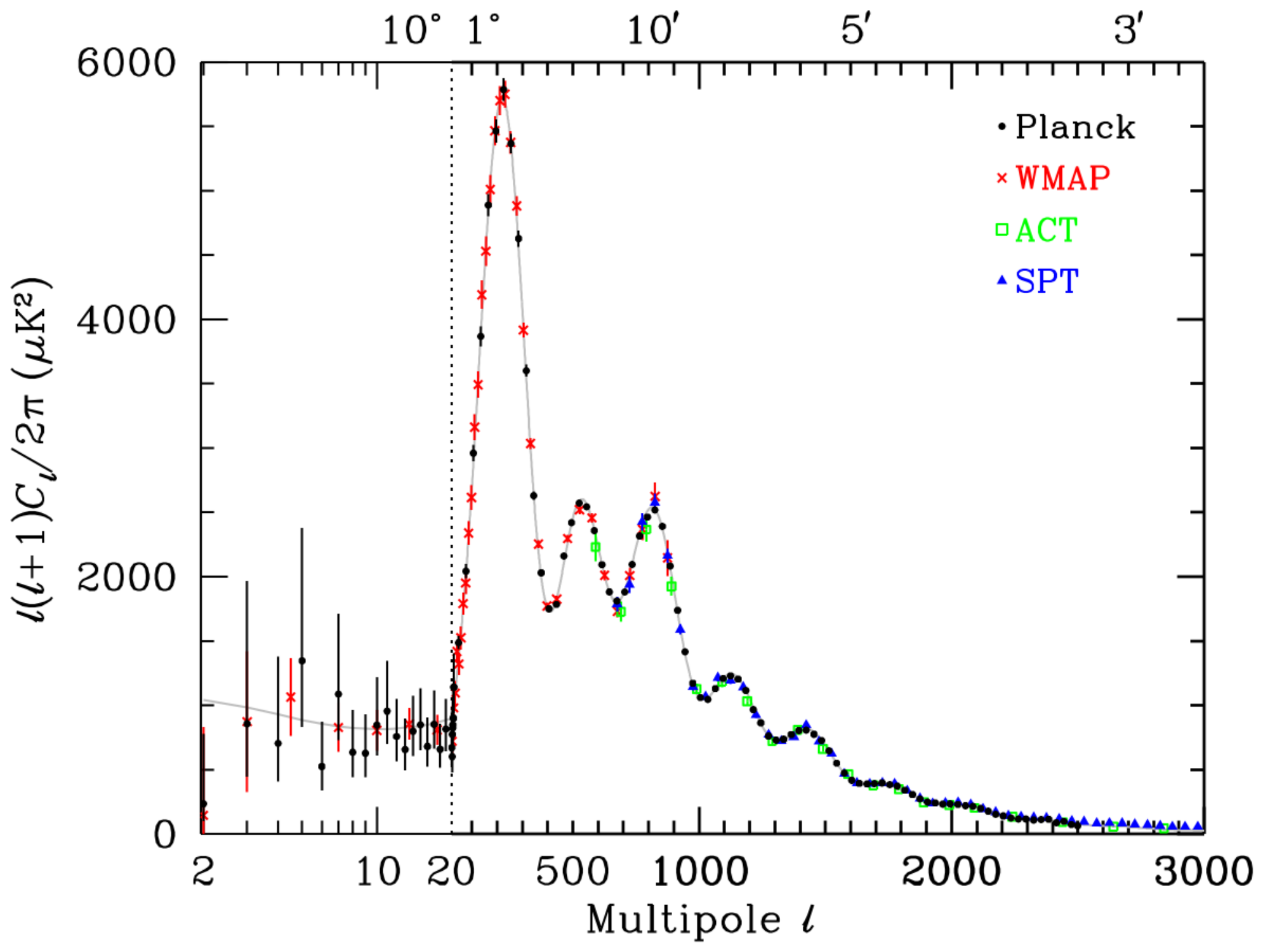}}
 \hfill
 \subfloat[]{\includegraphics[width=0.45\textwidth]{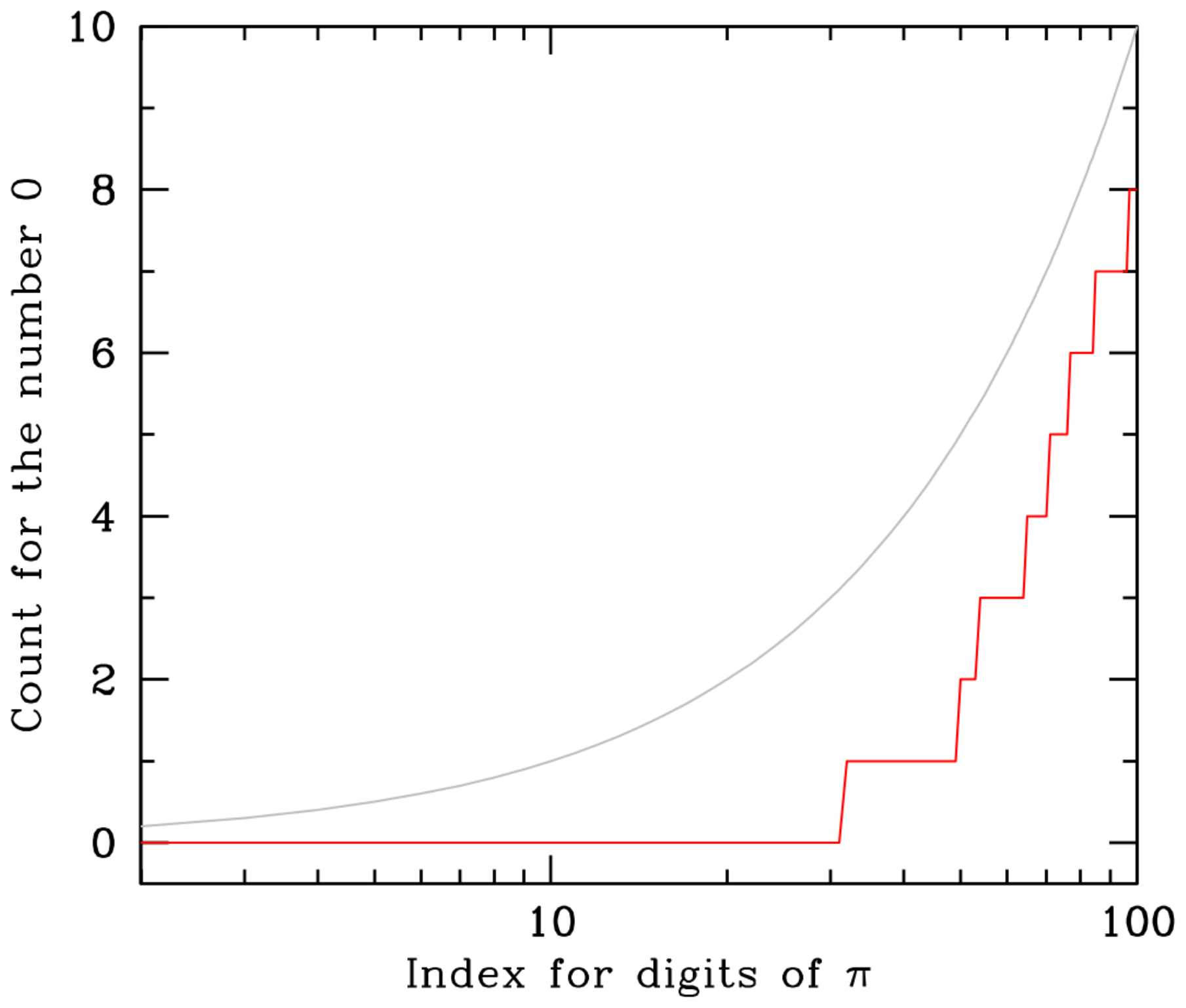}}
\end{center}
\caption{\label{fig:nodip} (a) Compilation of CMB power spectrum data from
{\it Planck}, {\it WMAP}, Atacama Cosmology Telescope \cite{ACT} and South Pole
Telescope \cite{SPT}.  As has become conventional, the lowest multipole part
is plotted logarithmically and the rest on a linear scale.  One can see that
over the wide range of multipoles that have now been well measured, the
deficit of power at $\ell=$20--30 really stands out.
(b). For $\pi$ we focus on the lowest integer, i.e.\ `0', and find that there
is a deficit in its abundance in the first digits (compared with the
expectation, shown by the grey line).  In fact the number 0 does
not occur at all until the 33rd digit.}
\end{figure*}

Another often-discussed anomaly is the lack of power in the $C_\ell$s at
low multipoles.  This manifests itself as a general lowering of power for
multipoles below about 40 and a specific `dip' that is noticed at
$\ell\simeq20$--30.  Such features would be buried at the lowest multipoles
if we were to plot $C_\ell$ versus linear $\ell$, but can easily be seen when
we carefully plot the lowest multipoles on a logarithmic scale, as shown in
Fig.~\ref{fig:nodip}(a).

This low-$\ell$ dip has a corresponding AAA in the digits of $\pi$,
manifesting itself in
the lack of the number `0'.  Fig.~\ref{fig:nodip}(b) shows the cumulative
count of 0 in the digits of $\pi$.  For the lowest digits there is an
obvious lack of 0s.  In fact the first 0 does not occur until the
33rd digit, the probability of which can
easily be estimated to be a few percent.  Hence this is interesting, but
perhaps not very remarkable in itself.  However, we find that consecutive
strings of the number 0 are also under-represented.  For example the
pattern `000000' is the only string of six consecutive numbers that does
{\it not\/} occur in the first million digits of $\pi$; in fact it does
not show up until digit 1{,}699{,}927.

Furthermore, if we consider the binary digits of $\pi$ we discover that there
are {\it many more\/} instances of 0 than 1 for the lowest digits.
In particular
61\,\% of the first 164 bits are 0, which is extreme enough to have
a probability of just 0.3\,\% (using the binomial distribution for 164 trials).
The fraction of occurrences of `0' stays well above 50\,\% for several hundred
more bits, showing that the binary digits also exhibit non-random features
that require an explanation.

\subsection{Hemispherical asymmetry}
\begin{figure*}[!htpb]
\begin{center}
 \subfloat[]{\includegraphics[width=0.5\textwidth]{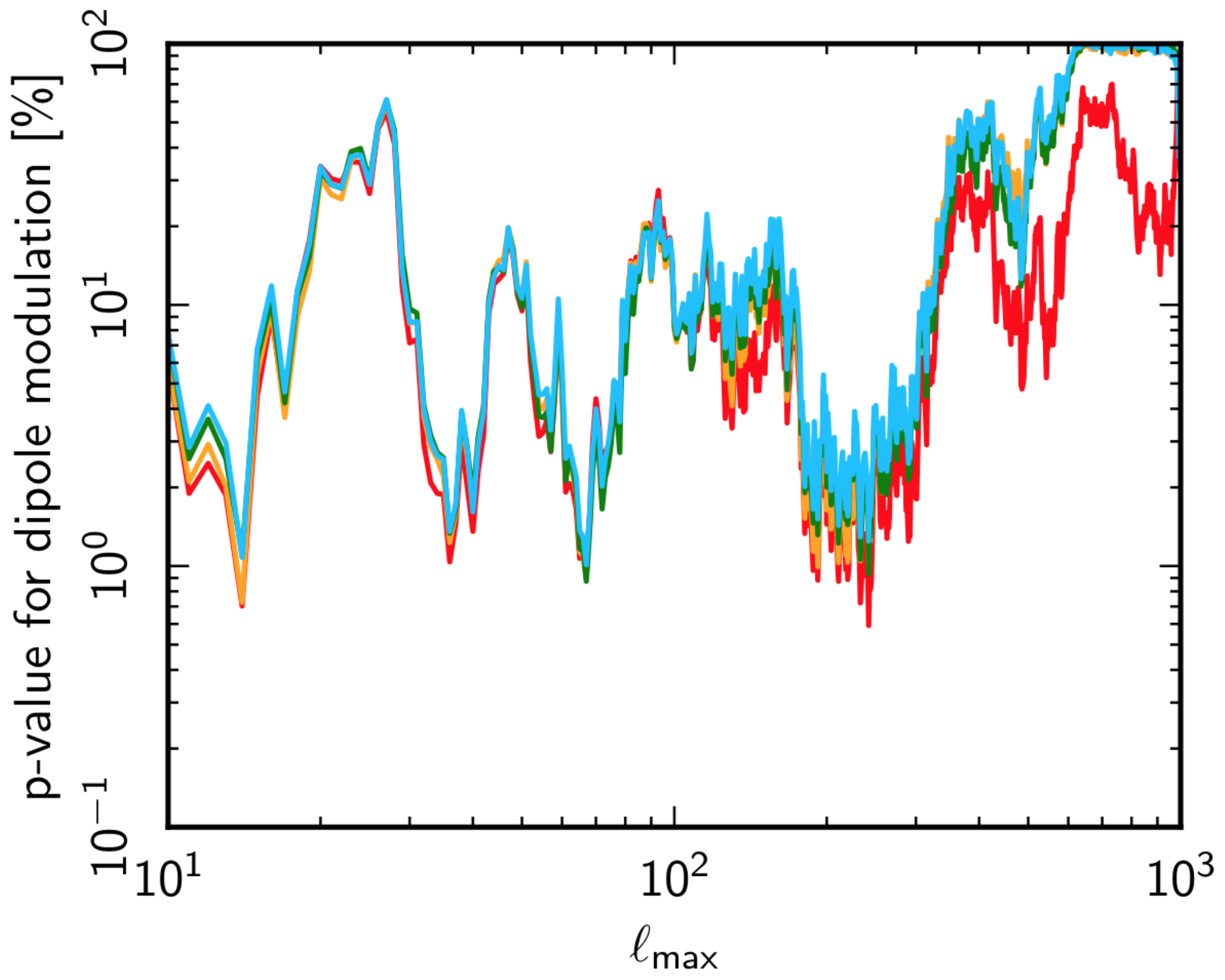}}
 \hfill
 \subfloat[]{\includegraphics[width=0.5\textwidth]{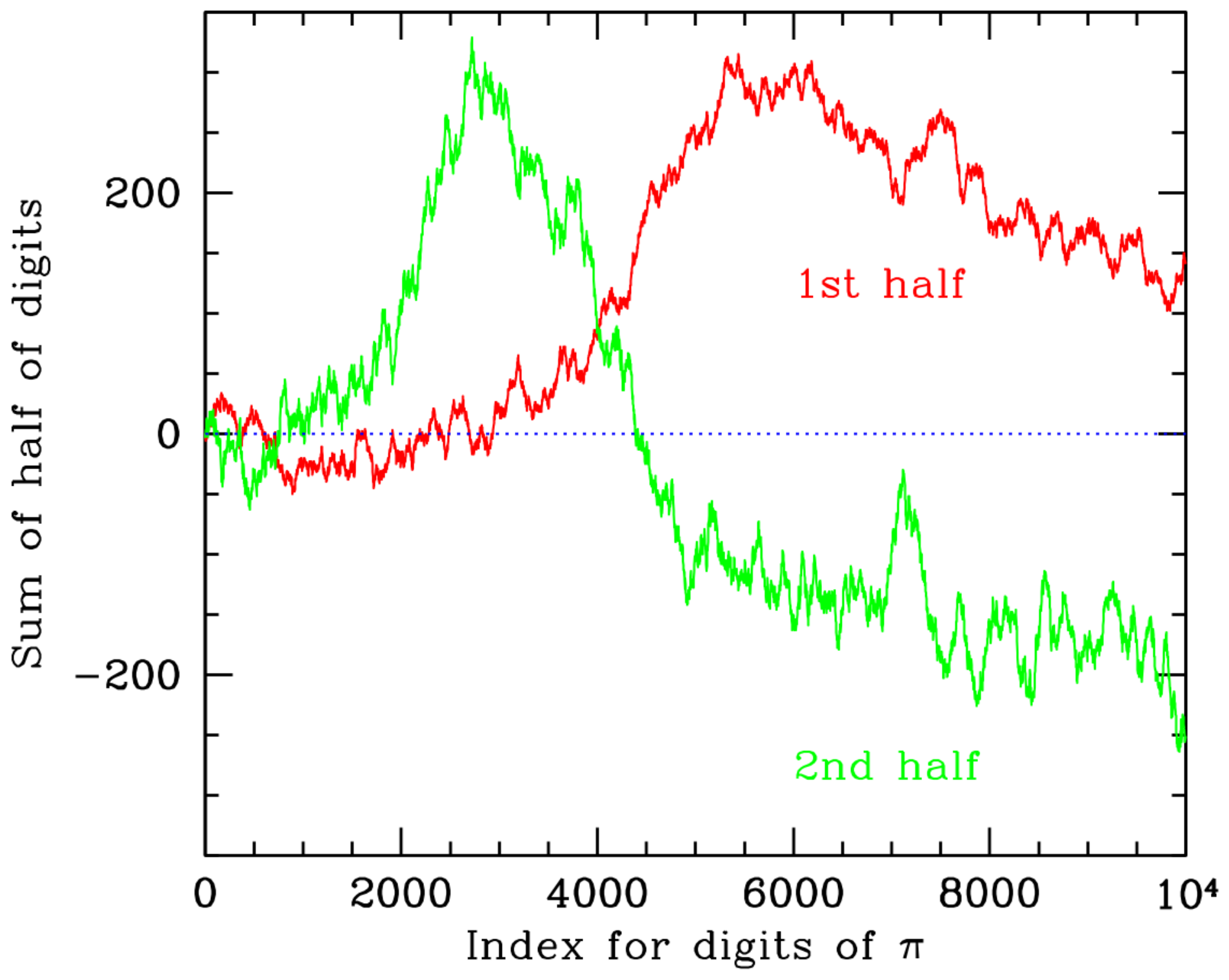}}
\end{center}
\caption{\label{fig:hemisphere} (a) On large angular scales the CMB sky has
more power in one hemisphere than the other, which can also be thought of as
dipole modulation of the sky.  Here (taken from Ref.~\cite{IandS}, and plotted
for four different foreground-separated CMB maps) we show
how the significance of this modulation varies with the maximum multipole
considered.  It is clear that the spike at $\ell\simeq65$ stands out
compared with all other scales.  The amplitude of the dipole modulation at
this scale is only
found in about 1\,\% of random simulations.  (b) If we take the digits of
$\pi$ out to some maximum digit and separately add the first half and
second half (after removing the average), we obtain the red and green lines,
respectively.  It is clear that the two halves of the digits behave in a
remarkably different way.}
\end{figure*}

Many studies have pointed out that the CMB sky can be split into two
hemispheres in which there is a dramatic difference in the power between each
half of the sky.  This can be studied equivalently by determining the
amplitude of the dipole modulation of the sky.  In Fig.~\ref{fig:hemisphere}(a)
we reproduce a plot of the probability of obtaining the observed dipole
amplitude in simulations of Gaussian skies (plotted here for each of the
four distinct foreground-separated CMB maps produced by the {\it Planck\/}
team, but these differences are not relevant here).  It is clear that the
scale $\ell_{\rm max}\simeq65$ is quite special, and yields a low probability
of about $1\,\%$.

We can examine an analogous statistic in the digits of $\pi$.  To do so we
consider the digits up to some maximum digit, and split them into the
first half and the second half.  By removing the average (i.e.\ the value
4.5 for the set of integers 0 to 9),
we calculate the sum of the two separate halves
of the digits, and plot this against the maximum digit in
Fig.~\ref{fig:hemisphere}(b).  The two halves of the digits of $\pi$ behave
in a dramatically different way, with an index around 2700 being the
preferred scale here, but the difference continuing out to the highest digits
examined.  This AAA thus demonstrates that the digits of $\pi$ have the same
sort of `lop-sidedness' as the CMB sky.

\subsection{Alignments}
\begin{figure*}[!htpb]
\begin{center}
 \subfloat[]{\includegraphics[width=0.575\textwidth]{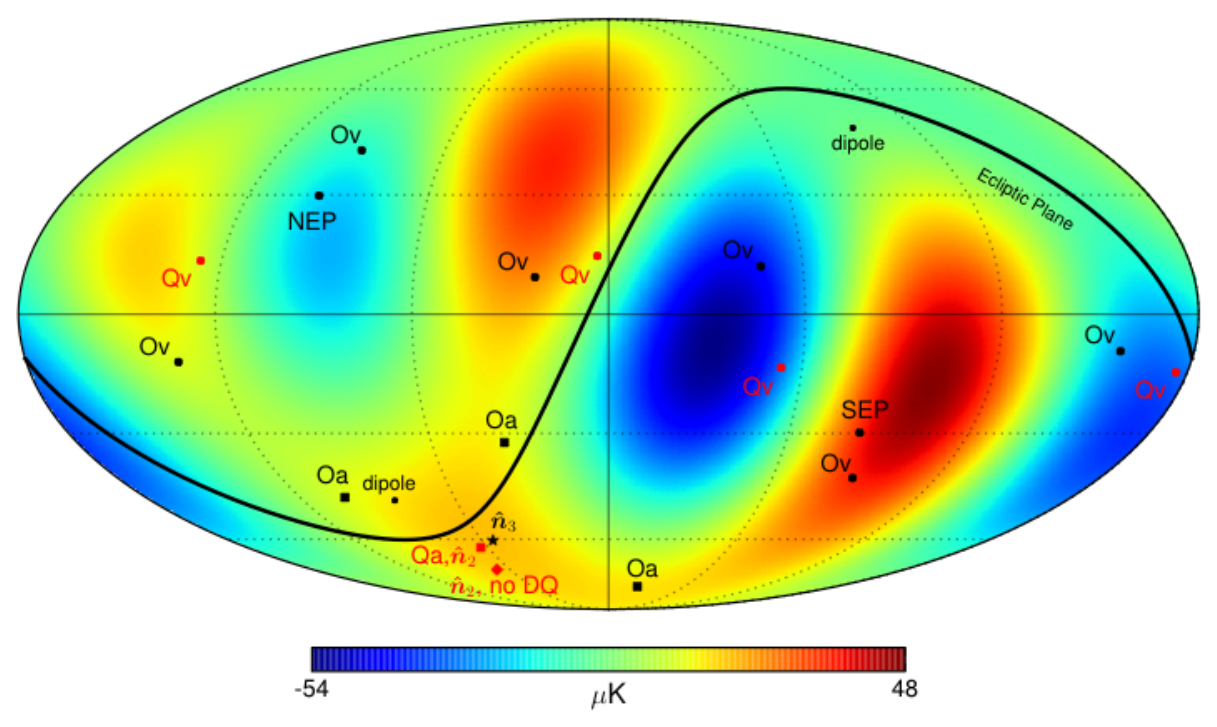}}
 \hfill
 \subfloat[]{\includegraphics[width=0.425\textwidth]{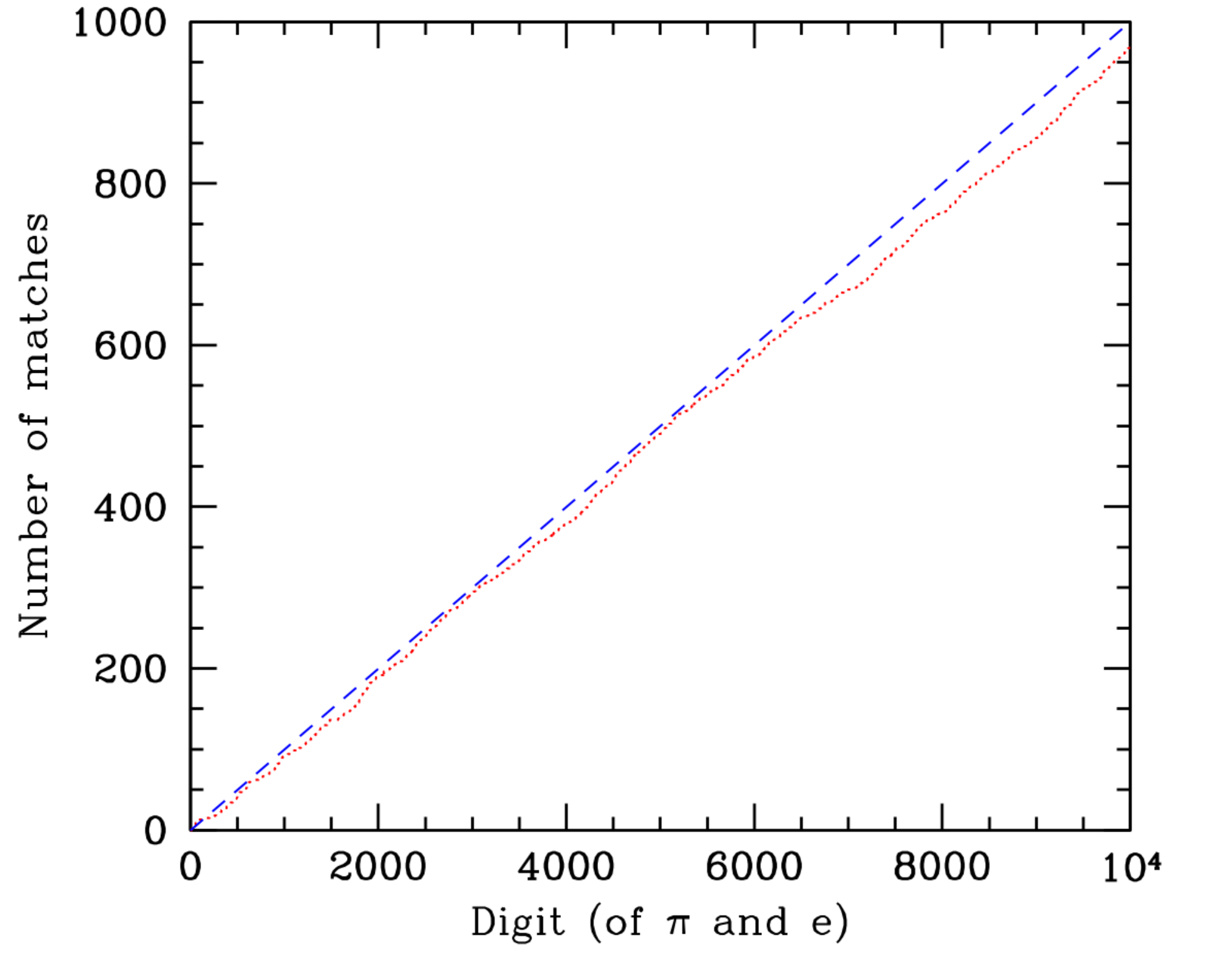}}
\end{center}
\caption{\label{fig:alignment} (a) Alignment of the dipole (D), quadrupole (Q)
and octupole (O) directions, taken from Ref.~\cite{improbable1}.  Since the dots
all appear in one small part of the sky, one can see that these special
directions are remarkably well aligned. (b) Anti-alignment of the digits of
$\pi$ and e.  If one compares these two numbers, digit by digit, it becomes
apparent that the fraction of matches found falls systematically short of
expectations for essentially all the digits investigated.}
\end{figure*}

Obviously the lowest-order CMB multipoles are special, and so should be
examined particularly closely.  It has been noticed that there is an `axis of
evil' \cite{axis}, in that the preferred directions of these lowest multipoles
are dramatically lined up with each other \cite{evil}.
The extraordinarily tight alignment of
the axes of the dipole, quadrupole and octupole are demonstrated in
Fig.~\ref{fig:alignment}(a).  This has been estimated to have a probability
of order 0.1\,\%.

We can search for an AAA alignment \cite{AAAA} in $\pi$ itself,
but in order to have
something to align it with we need to find another number.  The obvious
choice is `e', which occurs in physics and mathematics almost as often as
$\pi$.  Indeed both numbers occur in the famous Euler identity
$e^{i\pi}=-1$.  This may be the most compact and meaningful mathematical
expression found by humanity so far \cite{euler}, since it includes the
operations of subtraction, multiplication and exponentiation, as well as the
numbers $i$, 1 and 0 \cite{zero}, in addition to $\pi$ and $e$.  Another
reason to combine $e$ with $\pi$ is the obvious significance of these two
numbers for humans, since the number of pairs of chromosomes possessed by
homo sapiens is 23 \cite{birthday}, and this is the integer bracketed by
$\pi^e$ and $e^\pi$ \cite{pie}.

When we compare $\pi$ and $e$ digit by digit, we find that there are
systematically fewer matches than expected.  In the first 130 or so digits
there are quite a few matches (e.g.\ in the 13th, 17th and 18th digits),
occurring at a little more than the expect rate of 1 time in 10.  However,
after the 130th digit the fraction of matches is consistently low all the
way out to the 10{,}000th digit, as shown in Fig.~\ref{fig:alignment}(b).
One can estimate the significance of this result, e.g.\ using a
Kolmogorov-Smirnov test on the deviation, yielding a probability corresponding
to approximately a $6\,\sigma$ effect.  This is a remarkable result, showing
that although $e$ and $\pi$ {\it should\/} be composed of random digits, and
hence entirely uncorrelated, they are in fact {\it anti}-correlated.  There are
two obvious possibilities: this relationship could come from a subtle
connection through the Euler identity; or, more fundamentally, it could be
pointing to these digits being far from random, but actually contain meaningful
patterns \cite{million}.

\subsection{Odd/even asymmetry}
\begin{figure*}[!htpb]
\begin{center}
 \subfloat[]{\includegraphics[width=0.5\textwidth]{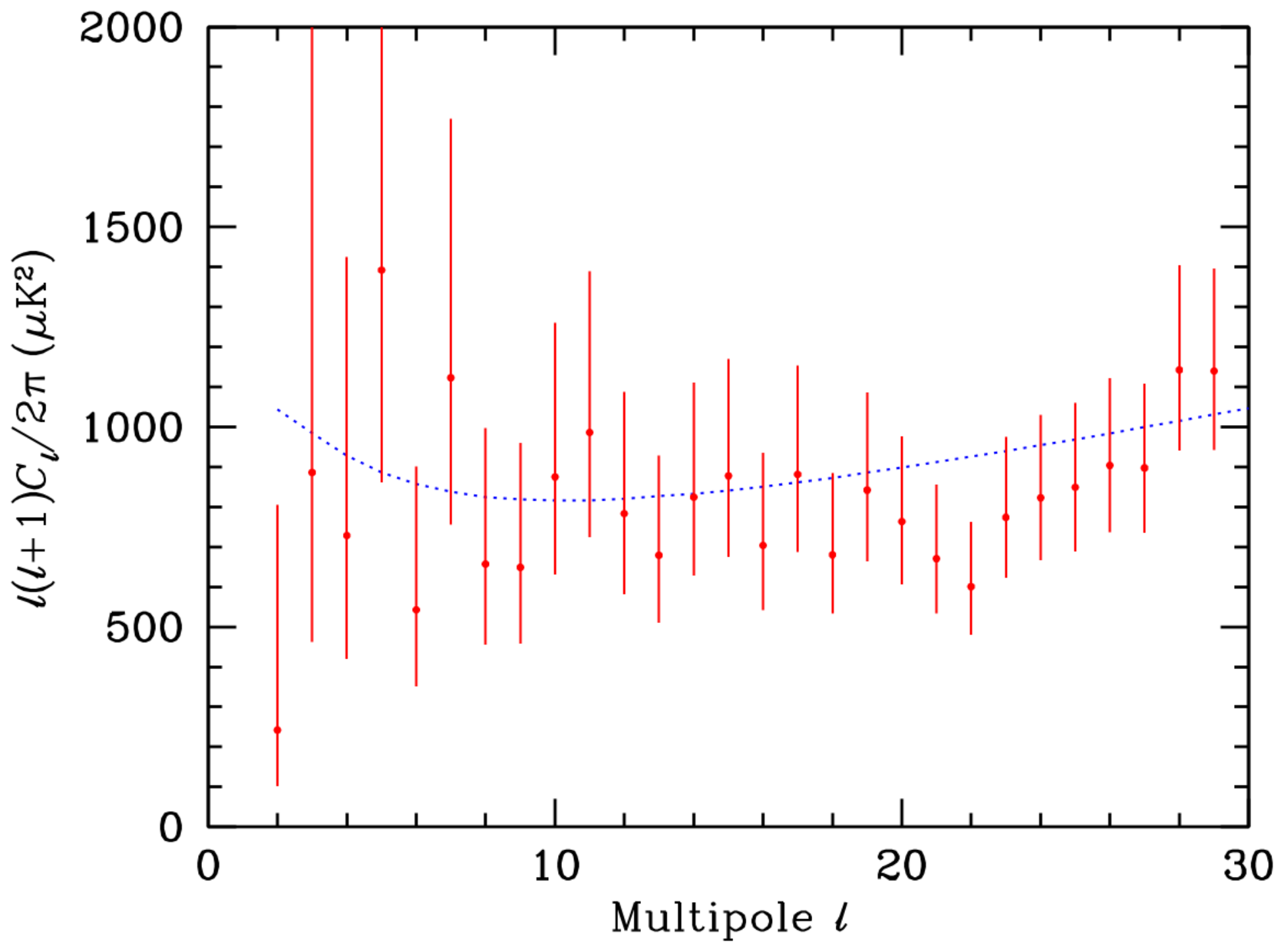}}
 \hfill
 \subfloat[]{\includegraphics[width=0.47\textwidth]{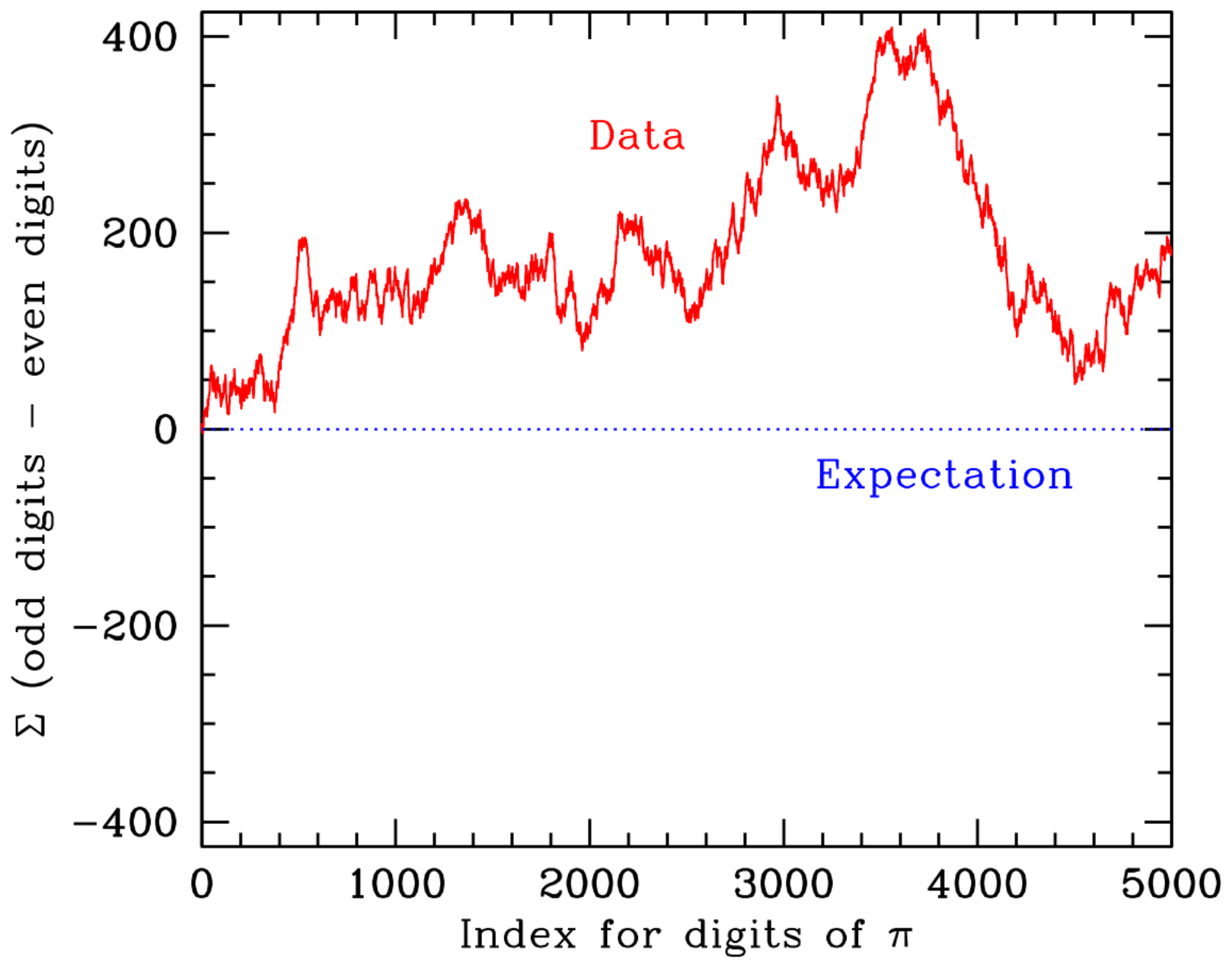}}
\end{center}
\caption{\label{fig:nosawtooth} (a) Power spectrum from {\it Planck\/}
data, showing the first 30 multipoles.  The `parity asymmetry' is evident
here, with a striking `saw-tooth' pattern of odd versus even multipoles.
(b) If we examine the digits of $\pi$ we find that the odd digits are
systematically higher than the even digits -- shown here by plotting the
cumulant of the sum of odd digits minus the sum of even digits.}
\end{figure*}

Another effect seen in the CMB power spectrum is that odd multipoles are
systematically higher than even multipoles, a property sometimes referred to
as `point-parity asymmetry'.  This can be seen in the {\it Planck\/} data
plotted in Fig.~\ref{fig:nosawtooth}(a), where there is a very striking
`sawtooth' pattern for odd versus even multipoles.

There is an analogy when we compare the odd and even digits of $\pi$.  If we
take the first digits of $\pi$ and split them into pairs, we find that the
first digit is sometimes bigger than the second one in the pair, but sometimes
the opposite is true.  For example in $3.14159\dots$, $3\,{>}\,1$ and
$4\,{>}\,1$, but $5\,{<}\,9$.  However, if we repeat this exercise,
then more often than not we find
that the odd digits are larger than the even digits.  This is shown in
Fig.~\ref{fig:nosawtooth}(b), where we have plotted the sum of the difference
between odd digits and even digits of $\pi$.  The probability of achieving
such a systematic difference over so many pairs of digits is exceedingly small.

One lesson here is that the odd digits of $\pi$ are generally bigger than the
even digits, and hence if you wanted, as a party trick, to learn {\it either\/}
the odd digits or even digits of $\pi$, you would be better to pick the odd
digits \cite{odddigits}.  But what could this `parity' AAA be telling us?

\subsection{Initials and words}
\begin{figure*}[!htpb]
\begin{center}
 \subfloat[]{\includegraphics[width=0.6\textwidth]{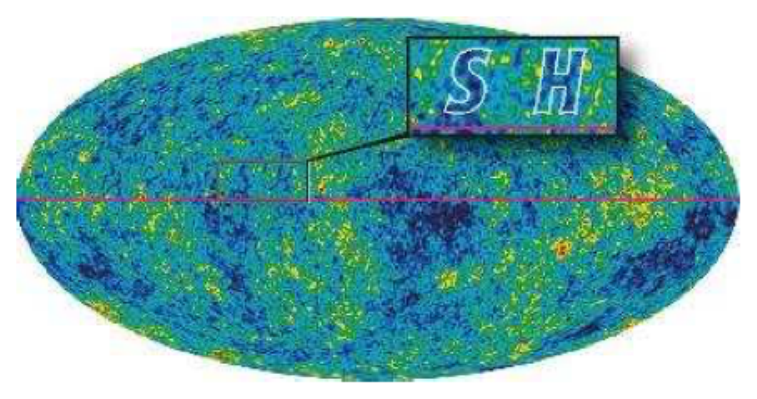}}
 \hfill
 \subfloat[]{\includegraphics[width=0.4\textwidth]{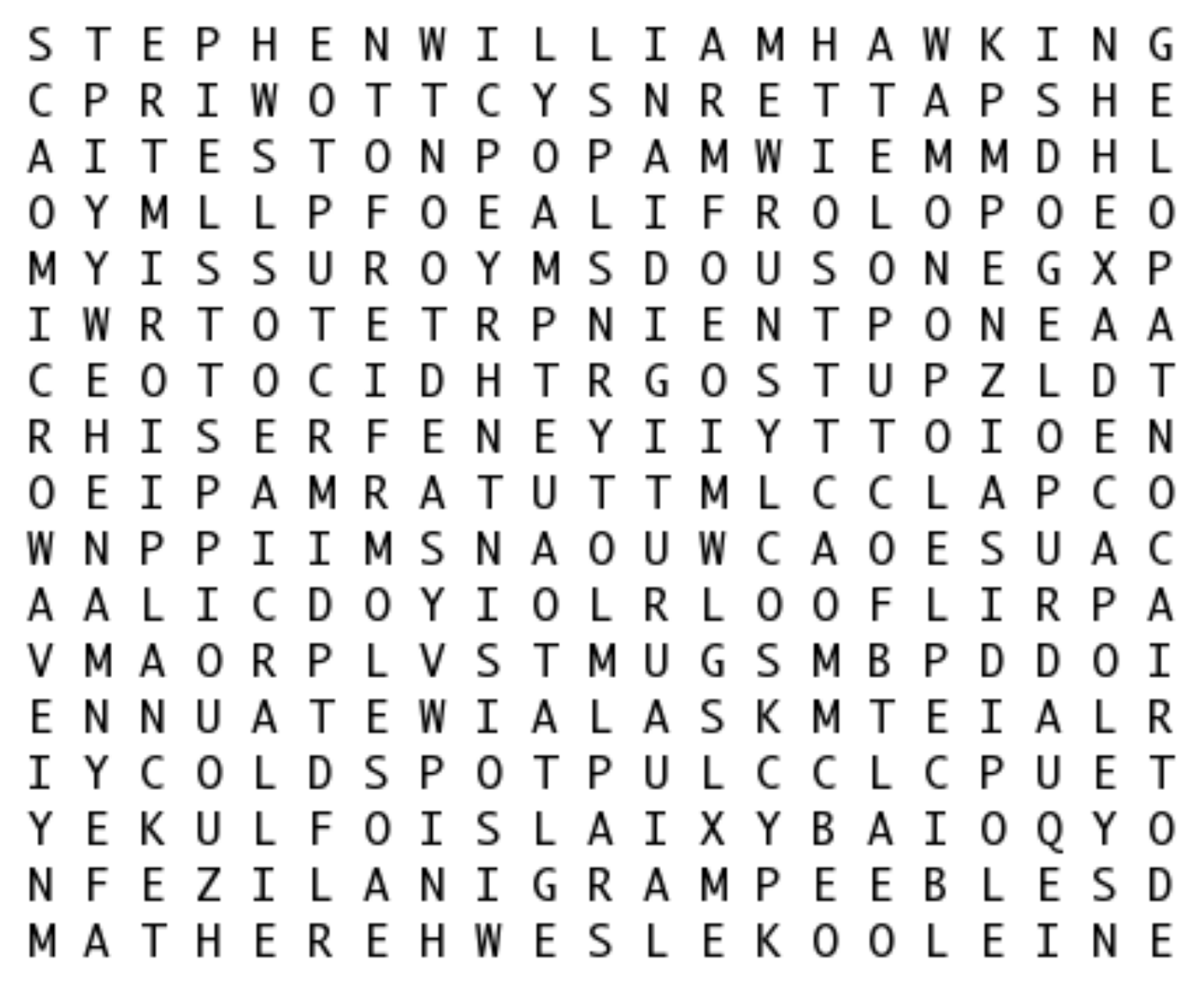}}
\end{center}
\caption{\label{fig:initials} (a) Indication of the initials `S.H.' that
appear on the CMB sky (taken from Ref.~\cite{WMAPA}).
(b) When we translate the digits of $\pi$ into letters, we can start to
see messages that are more unusual than mere initials.}
\end{figure*}

It was pointed out by the {\it WMAP\/} team \cite{WMAP} that one can find
Stephen Hawking's initials in the CMB sky, as shown in
Fig.~\ref{fig:initials}(a).  These same features, clearly being two roman
characters in roughly the same size, font and orientation, are also present in
the {\it Planck\/} data.  We are not aware of any systematic search for
all combinations of letters, with arbitrary position, font, size and
orientation.  But presumably such a search would turn up many other interesting
examples of writing on the `cosmic billboard'.

Words can also be searched for in the digits of $\pi$.  In order to do that one
needs to select a scheme for converting the digits into letters.  We are aware
of four separate methods to do this.  Firstly one could write the digits of
$\pi$ in base 36, so that after `0, 1, \dots 9' we have
`A, B, C, \dots', including
the whole of the alphabet.  Secondly one could just take base 26 and map it
on to the alphabet (or base 27, to include a space) \cite{base26}; using this
method we find that the word `odd' occurs considerably more often than the word
`even', which is just like the CMB point-parity asymmetry.
Thirdly one could determine $\pi$ in binary and then use groups
of five binary digits to map to the alphabet (plus a few punctuation marks)
\cite{bailey}; in this scheme in fact the word `anomaly' appears at binary
digit 1{,}769{,}000{,}467. Lastly one can use the more complicated
transformation that we adopted to generate Fig.~\ref{fig:initials}(b).

One sees in this figure that it is possible to find much more
noteworthy patterns
than the mere pair of initials found in the CMB sky.  For example, across
the top of the plot we can just about discern the middle initial for `S.H.'
And the right-hand half of the figure contains several aligned multipoles.
For readers interested in further studying this `word search' diagram, we
have provided details of our own investigations in an Appendix.
It is obvious that for this AAA, it is relatively easy to find features in the
digits of $\pi$ that are substantially more anomalous than what we see in the
CMB.

\section{Conclusions}

The CMB sky contains a combination of several independent
anomalies, with a joint
probability that makes what we observe extremely unlikely in the standard
model, hence indicating evidence for several kinds of new physics.  But perhaps
even more amazingly, there are similar kinds of feature in the digits of $\pi$
and these are every bit as significant.

The large-scale CMB sky presents a fixed pattern to us, and hence if one
finds anomalies in this pattern, there is no more sky to use to confirm or
improve the significance of what was initially found.  The situation with the
digits of $\pi$ is analogous, since any anomalies discovered cannot be
corroborated by performing experiments on these fixed digits \cite{Pi}.

Some have argued that such features in the digits of $\pi$ need to be
considered in the light of a posteriori statistics, or in other words that one
should consider the multiplicity of other tests that could have been carried
out, to marginalise over similar features that {\it might\/} have been found,
using what is sometimes called the `look-elsewhere effect' \cite{fishing}.
Others have argued that we should look elsewhere for an explanation, that
resorting to such statistical sleight-of-hand is the last bastion of
scoundrels, because it can be used to torpedo {\it any\/} apparently
${\simeq}\,3\,\sigma$ effect \cite{mix}.

Be that as it may, this issue about overestimating the statistical
significance of the anomalies \cite{lies} surely cannot apply to the many
dramatic features seen on the large-scale CMB sky.  And hence we must also
accept the importance of all the structures that also exist
within the digits of $\pi$.

What could these aberrations be telling us?  And how are they related to
other apparent coincidences \cite{us} in nature?  If we are seeing anomalies
in the decimal digits of $\pi$, then presumably there is something special
about base 10.  And hence we are indeed created in the image of the creator of
not only the entire physical Universe, but also the reality of mathematics.
Probably there is an anthropic argument that we could only exist in a
reality in which mathematics and physics contain these signatures.

No doubt further study of these joint anomalies will give us insight into
other dimensions and universes that are otherwise unobservable.  Indeed the
`Pi theorem' of Buckingham \cite{pitheorem} is a constraint on the number
of dimensions within physics.  Can we conceive of $\pi$ being different in
these other dimensions?  And does this teach us about the nature of the
relationship between physics, mathematics and theology \cite{barrow}?
Another connection is that the CMB temperature was once equal to $\pi\,$K, and
that epoch corresponded to a redshift of $z\,{=}\,0.153$, about 2\,billion
years in the past \cite{CMBtemp} -- did something happen on
Earth around this time that connected the development of life
with the background radiation and the digits of $\pi$?

As a final remark, we add this utilitarian view -- if the AAAs mean that the
CMB information is somehow {\it already\/} encoded in $\pi$, then perhaps in
future we can avoid all the fuss and bother of building real CMB experiments,
minimising systematic effects while operating them, painstakingly analysing
the data, and debating the statistical interpretation of the results -- and
instead simply look more carefully at the digits of $\pi$,
or in any other random string of digits \cite{further}.


\smallskip


\appendix
\section{Searching for words}
Fig.~\ref{fig:initials}(b) presented some of the digits of $\pi$ transformed
into the normal letters of the alphabet.  Precisely which digits these are,
and how we performed the translation, are uninteresting details -- we guarantee
that this set occurs {\it somewhere\/} in $\pi$.

We have performed extensive analysis of this set of letters, searching for
meaningful words.  We believe we can find all of the following:
stephen william hawking;
peebles, penzias, wilson, mather, smoot;
euler, feynman, gauss;
pi (several times), pie (twice), digit;
cmb, cosmic, microwave, background, anisotropy;
multipole, monopole, dipole, quadrupole, octupole, hexadecapole,
 dotriacontapole;
act, bicep, cobe, planck, spt, wmap;
cold spot, hot spot, low-l dip, alignment, parity, anomaly, hemispheric,
 asymmetry, stripe, coldest, deviations;
a posteriori, look elsewhere, marginalize, fluke;
forty-two;
ali frolop;
april fool.
Since the {\it lack\/} of a deviation may carry the most information,
when all these words have been found, the remaining letters may spell out the
ultimate question \cite{douglas}.


\begin{thebibliography}{9}
\bibitem{sesame} It has been said, for example, that if the {\it Planck\/}
 results were an episode of Sesame Street, they would be brought to you by the
 words `tension' and `anomaly'.
\bibitem{flip} There are many stories about why $\pi$ is used as the symbol
 for this number, for example that the reflection of {\sf 3.14} looks like
 \flip{{\sf 3.14}}.
\bibitem{recite} Reciting the digits of rational numbers (like `3' for
 example) would be much more straightforward.
\bibitem{piday} Although only really in North America, and not, for example,
 in Europe, since there is no 31st of April.
\bibitem{pilish} Jeans is meant to have composed this pilish
 sentence corresponding to the first 15 digits of $\pi$: `How I need a drink,
 alcoholic in nature, after the heavy lectures involving quantum mechanics!'
 For the topic of this paper we offer this clumsy effort: `And a mark I found
 entwining pi, belies skies!  CMB alike isotropy anomalies dwindle\dots'.
\bibitem{chart} Images of the microwave sky of this sort are usually called
 `CMB maps', while diagrams showing the digits of $\pi$ are referred to as
 `$\pi$ charts'.
\bibitem{coldest} Moreover, most of the rest of the sky is a dramatic
 $150\,\mu$K hotter.
\bibitem{Feynman} After Richard Feynman, who is sometimes credited with the
 story that if you could learn $\pi$ to 762 digits, you could say `nine, nine,
 nine, nine, nine, nine, and so on'.  However, there is no documented evidence
 that Feynman ever said this.  It is probably an example of Stigler's Law of
 Eponymy.
\bibitem{tau} Emphasized by Michael Hartl as part of a claim that $2\pi$ is
 more than twice as good as $\pi$.  See {\tt tauday.com}.
\bibitem{evil} Although we don't understand what is `evil' about it, and
 instead prefer the term `shaft of light'.
\bibitem{euler} Perhaps even more important than some of the patterns we have
 found in the digits of $\pi$.  We leave as a future exercise the search for
 Euler's identity encoded into the digits of $\pi$ and $e$ themselves.
\bibitem{zero} Although it is a bit harder to see the zero than the other
 numbers.
\bibitem{pie} Furthermore, everyone knows that if you combine pi and e you
 get pie, which is usually a good thing.
\bibitem{odddigits} Of course you could easily trick people by claiming to
 know all the odd digits of $\pi$, and then simply reciting `1, 3, 5, 7 and 9'.
\bibitem{bailey} This is the method used at David Bailey's site
 {\tt pi.nersc.gov}, where one can search for one's name in the first 4 billion
 binary digits of $\pi$.  The fact that the word `hot' occurs in these digits
 dramatically earlier than the word `cold' is further demonstration that $\pi$
 prefers hot spots rather than cold spots.
\bibitem{Pi} However, for the CMB, it may be possible to search for related
 anomalies in the polarization data, $\Pi$.
\bibitem{mix} And to mix metaphors where no one has gone before.
\bibitem{lies} Mark Twain would no doubt have quipped that this is a debate
 about `anomalies, damned anomalies, and statistics'.
\bibitem{CMBtemp} C.\ Uhlenberg, private communication; this may be consistent
 with the time of the first binary black hole merger detected by LIGO.  It may
 also be worth noting that at the last-scattering epoch the CMB temperature
 was $\pi^7\,$K.
\bibitem{further} Comments on this paper should be sent to the lead
 author, Dr.~Frolop.  We would like to thank CAASTRO for their hospitality
 when the last parts of this paper were being completed -- one of us spent
 time at the University of Queensland and the other at Mount Stromlo
 Observatory, and both of us enjoyed our time in Australia, where they
 really like pies.
\end{thebibliography}
\end{document}